\documentstyle[letters]{mn}
\input epsf
\title[Irrotational binary neutron star systems]
{Stationary states of irrotational binary neutron star systems 
and their evolution due to gravitational wave emission}
\author[K. Ury\=u and Y. Eriguchi]
{K. Ury\=u$^{1,2,3}$ and Y. Eriguchi$^3$\\
$^1$International Center for Theoretical Physics, Strada Costiera 11,
34100 Trieste, Italy\\
$^2$SISSA, Via Beirut 2/4, 34013 Trieste, Italy\\
$^3$Department of Earth Science and Astronomy, 
Graduate School of Arts and Sciences, 
University of Tokyo\\ Komaba, Meguro, Tokyo, Japan
}
\begin{document}

\maketitle
%
\begin{abstract}
We have succeeded in obtaining {\it exact} configurations of irrotational 
binary systems for compressible (polytropic) equations of state.  Our 
models correspond to binaries of equal mass neutron star systems in the 
viscosity free limit.  By using the obtained sequences of stationary states, 
the evolution of binary systems of irrotational neutron stars due to 
gravitational wave emission has been examined. For inviscid binary systems the 
spin angular velocity of each component in a detached phase is smaller than 
the orbital angular velocity at a contact phase. Thus the irrotational 
approximation during evolution of binary neutron stars due to gravitational 
wave emission can be justified. Our computational results show that the binary 
will {\it never} reach a dynamically unstable state before a contact phase 
even for rather stiff polytropes with the index $N \ga 0.7$, as the separation 
of two components decreases due to gravitational wave emission.  This 
conclusion is quantitatively different from that of Lai, Rasio \& Shapiro 
who employed {\it approximate} solutions for polytropic binary systems.
\end{abstract}

\begin{keywords}
binaries:close -- hydrodynamics -- instabilities -- stars:neutron -- 
stars:rotation
\end{keywords}

\section{Introduction and summary}

Coalescing stages of binary neutron star systems due to gravitational 
wave emission have been considered to be one of the most important 
targets of the advanced gravitational wave detectors (LIGO/VIRGO/GEO/TAMA, 
see e.g. Abramovici et al. 1992 and Thorne 1994).  Observations of 
final stages of binary neutron star systems will provide us a large
amount of new information about macroscopic quantities such as the mass
and the spin of neutron stars as well as about microscopic characters 
such as the viscosity and the equation of state of neutron star matter 
\cite{cu93}.  

Kochanek \shortcite{ko92} and Bildsten \& Cutler \shortcite{bc92} pointed 
out that viscosity in neutron stars may not be so effective as to 
synchronize the spin and the orbital angular velocity on a time scale of 
evolution due to gravitational wave emission.  For such inviscid fluids, 
Ertel's theorem holds: the ratio of the vorticity vector $\bzeta_0$ in the 
inertial frame to the density $\rho$ of a fluid element, $\bzeta_0/\rho$, 
is conserved even under the existence of a potential force such as the 
gravitational radiation reaction \cite{mi74}.  The vorticity vector in the 
inertial frame is defined as 
\begin{equation}
\bzeta_0({\bf x}) \equiv \nabla\times{\bf v}({\bf x}) \ \ , 
\end{equation}
where ${\bf x}$ and ${\bf v}({\bf x})$ are the position vector of the fluid 
element and the velocity field seen from the inertial frame, respectively.  
Since the spin angular velocity of each component of a detached binary system
is expected to be at least 50 times smaller than the orbital angular velocity 
at an almost contact state of the binary system (see the references above), 
we may consider that real close binaries have {\it irrotational} 
configurations. In other words, the vorticity of each component star seen 
from the inertial frame of reference can be neglected, i.e. 
$\bzeta_0 \, = \, 0$.  

It is not, however, an easy task to construct consistent models of stationary 
configurations of compressible stars such as binary systems with {\it 
arbitrary spins} (see e.g. Ury\=u \& Eriguchi~(1996) and references therein).  
Therefore compressible binary configurations in equilibrium states have been 
investigated only for {\it synchronized} binary systems in Newtonian 
gravity (e.g. Hachisu \& Eriguchi 1984; Hachisu 1986) or in post-Newtonian 
gravity (Shibata 1994, 1997) up to highly deformed configurations.  
As for configurations whose spins are different from the orbital angular 
velocities, there are only approximate solutions by Lai, Rasio \& 
Shapiro~(1993a, 1994a (hereafter LRS1), 1994b (LRS2)).  They employed
triaxial ellipsoidal polytropes for deformed binary states in Newtonian 
gravity and discussed evolutions of binary neutron stars.  Since real 
configurations are no more ellipsoidal when the binary comes near to 
a contact phase, their results cannot give quantitatively correct values 
even in Newtonian gravity. 

Recently we have formulated a scheme to compute irrotational binary 
configurations composed of two compressible stars with equal mass 
and developed a numerical code to solve exact configurations for such
binary systems in Newtonian gravity \cite{ue97b}.  Using this newly 
developed code, we have constructed stationary sequences of irrotational binary
systems. Obtained models have been used to investigate evolutions of binary 
systems due to gravitational wave emission.  Our computational results show 
that dynamical instability will {\it not} set in before a contact phase for 
polytropes with the polytropic index $N \ga 0.7$ as the separation of two 
components decreases.  For polytropes with the index in this range, dynamical 
instability will occur on a so-called `dumbbell' sequence \cite{eh85} for 
irrotational self-gravitating fluids.  This result is different from that 
of Lai, Rasio \& Shapiro~(1993a, 1994a, 1994b) who concluded that dynamical 
instability sets in on polytropic binary sequences with $N \la 1.2$ by
using ellipsoidal approximations of deformed configurations.  
In this Letter we will briefly summarize our new result and discuss its 
physical relevance in evolution of binary neutron star systems.  

\section{Formulation of the problem}

We consider {\it stationary} structures of polytropic binary stars without 
viscosity in the rotating frame.  We assume that the binary is composed of 
equal mass components whose vorticities equal to zero in the inertial frame
of reference.  For such irrotational cases, we can introduce the velocity 
potential $\Phi({\bf x})$ as follows:
\begin{equation}
{\bf v} = \nabla \, \Phi \, ,
\end{equation}
where ${\bf v}$ is the velocity vector in the inertial frame.  
The Euler equation of fluids can be integrated to the generalized 
Bernoulli's equation as follows in the rotating frame of reference: 
\begin{equation} \label{bern1}
{\partial \, \Phi \over \partial \, t} \, - \,({\bf \Omega} \times {\bf r})
\cdot \nabla \, \Phi \, + \, {1 \over 2} \mid \nabla \, \Phi \mid^2 \, + \,
\int{dp \over \rho} \, + \, \phi \, = \, f(t), 
\end{equation}
where $p$, $\phi$, ${\bf \Omega}$ and ${\bf r}$ are the pressure, 
the gravitational potential, the orbital angular velocity 
vector of the binary which is identical to the angular velocity vector of 
the rotating frame relative to the inertial frame and the position vector of 
the fluid element of the star from the rotational center, respectively, and 
$f(t)$ is an arbitrary function of time.  We note that the steady velocity 
${\bf u}$ in the rotating frame of reference is related to the velocity 
${\bf v}$ in the inertial frame as follows:
\begin{equation}
{\bf u} = {\bf v}\,-\,{\bf \Omega}\times{\bf r}\, .
\end{equation}

In actual computations we use 
the polytropic relation and the Emden function defined as: 
\begin{equation} \label{poly}
p = K \rho^{1+1/N} = K \Theta^{N+1},
\end{equation}
where $\Theta$ is the Emden function and $K$ is a constant.  Since the 
configuration of the binary is assumed to be stationary in the rotating 
frame, we can set 
\begin{equation} \label{stat1}
{\partial \, \Phi \over \partial \, t} \, \equiv \, 0 \quad{\rm and }\quad 
f(t) \, = \, C \, = \, constant \, . 
\end{equation}
Then equation (\ref{bern1}) becomes as
\begin{equation} \label{bern2}
\Theta \, = \, {1 \over K\,(N+1)}
\, [ \, ({\bf \Omega} \times {\bf r}) \cdot \nabla \, \Phi \, - \,
{1 \over 2} \mid \nabla \, \Phi \mid^2 \, - \, \phi \, + \, C \, ] \,.
\end{equation}

The equation of continuity is expressed by using the velocity potential 
as follows also in the rotating frame:
\begin{equation} \label{conti}
\nabla^2 \Phi \, = \, N ({\bf \Omega} \times {\bf r} \, - \, 
\nabla \, \Phi) \cdot {\nabla \, \Theta \over \Theta}, 
\end{equation}
where the assumption of a stationary state in this frame has been taken 
into account as follows:
\begin{equation} \label{stat2}
{\partial \, \rho \over \partial \, t} \, \equiv \, 0 \, . 
\end{equation}

After substituting the following integral expression for the Newtonian 
gravitational potential 
\begin{equation} \label{pote}
  \phi({\bf r}) \, = \, - \, G 
  \int_V {\rho({\bf r}^{'}) \over \mid {\bf r}-{\bf r}^{'} \mid } 
  d^3 {\bf r}^{'},
\end{equation}
into equation (\ref{bern2}), where integration is performed over the whole 
stellar interior $V$, there remain two unknown physical variables in the 
basic equations, i.e. the Emden function $\Theta$ and the velocity 
potential $\Phi$.  We can regard equation (\ref{bern2}) as the equation
for the variable $\Theta$ and equation (\ref{conti}) as that for the 
variable $\Phi$.  The boundary conditions for these two variables are as 
follows: 
\begin{eqnarray} \label{bcon}
(\nabla\,\Phi \,-\,{\bf \Omega}\times{\bf r}) \cdot {\bf n} \, = \, 0, 
\quad & {\rm on\ the\ stellar\ surface,} \\
\Theta \, = \, 0, \quad & {\rm on\ the\ stellar\ surface.}
\end{eqnarray}
These basic equations are transformed into the surface fitted coordinate system
(Ury\=u \& Eriguchi 1996) and solved iteratively by using the self-consistent 
field method (Hachisu 1986).  The detailed numerical method will be explained 
in the forthcoming paper \cite{ue97b}.  

\begin{figure}
\epsfbox{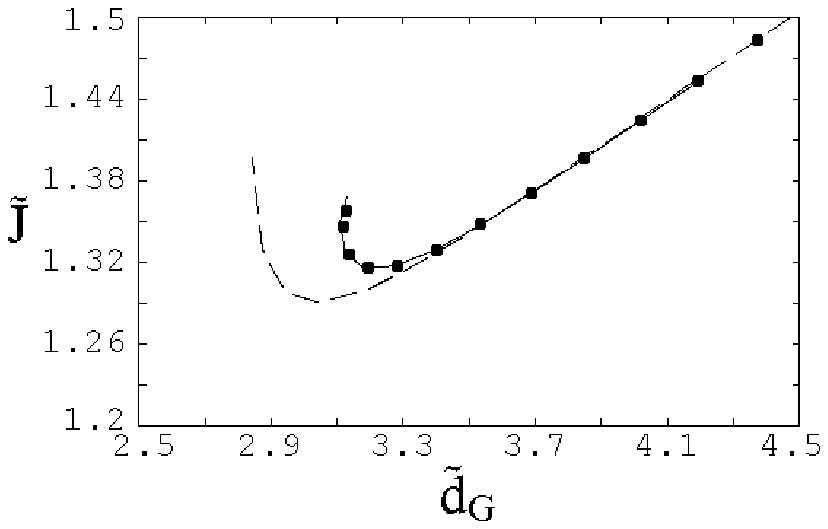}
\vspace{0.3cm}
\epsfbox{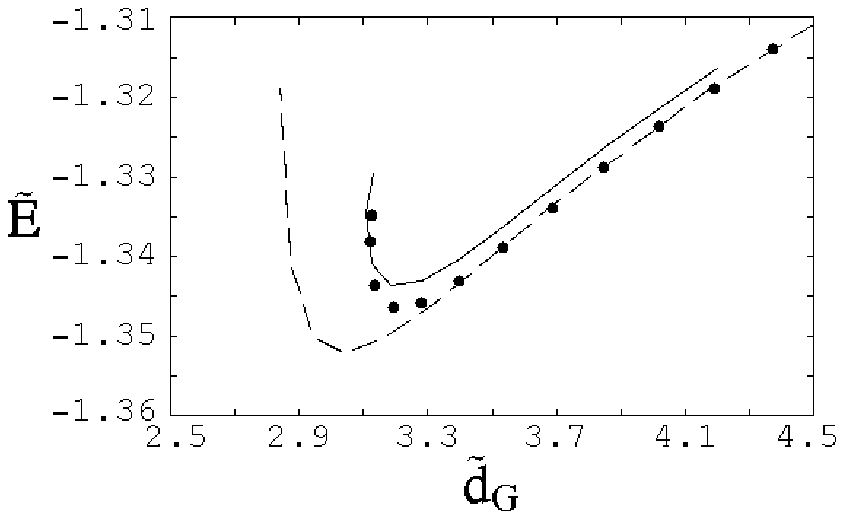}
\vspace{0.3cm}
\epsfbox{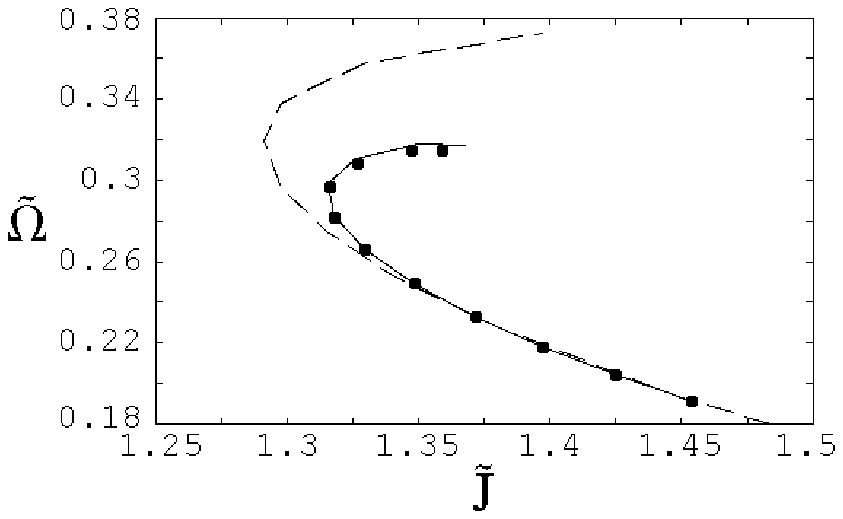}
\vspace{0.1cm}
\caption{Physical quantities of stationary sequences of irrotational binary 
states for $N=0$ incompressible fluid stars.  (a) Total angular momentum 
as a function of a binary separation. (b) Total energy as a function of 
a binary separation. (c) Orbital angular velocity as a function of the
total angular momentum.  Dashed and solid curves show the results of LRS1 and 
our present results of irrotational binary stars, respectively.  
Dots show the results computed by using different numerical method 
(Ury\=u \& Eriguchi 1997a).  See text about the normalization factors 
for each quantity.}
\label{fig-N=0}
\end{figure}

\begin{figure}
\epsfbox{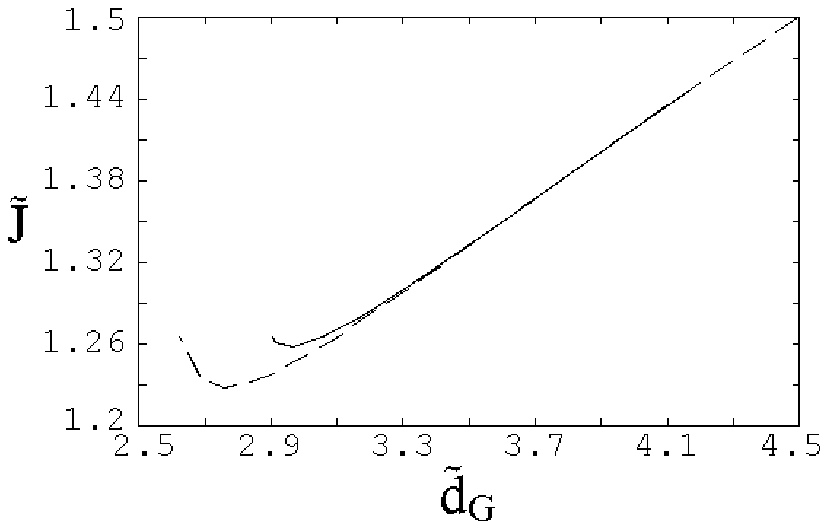}
\vspace{0.3cm}
\epsfbox{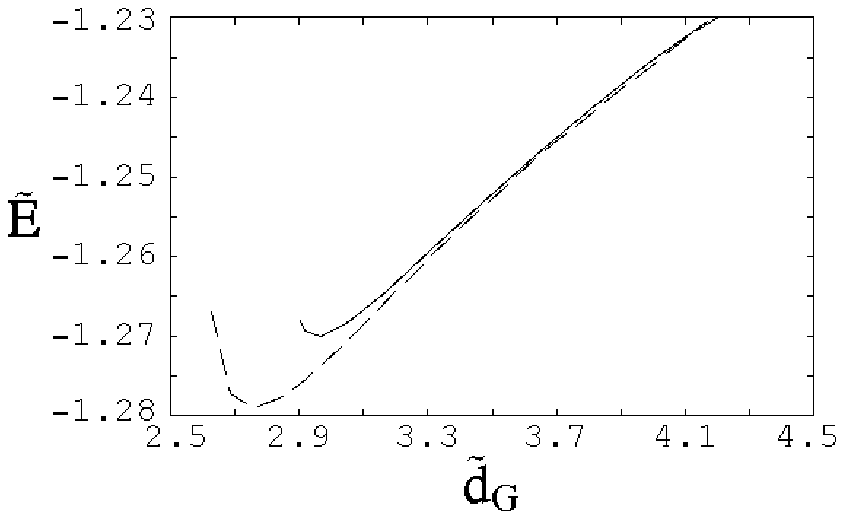}
\vspace{0.3cm}
\epsfbox{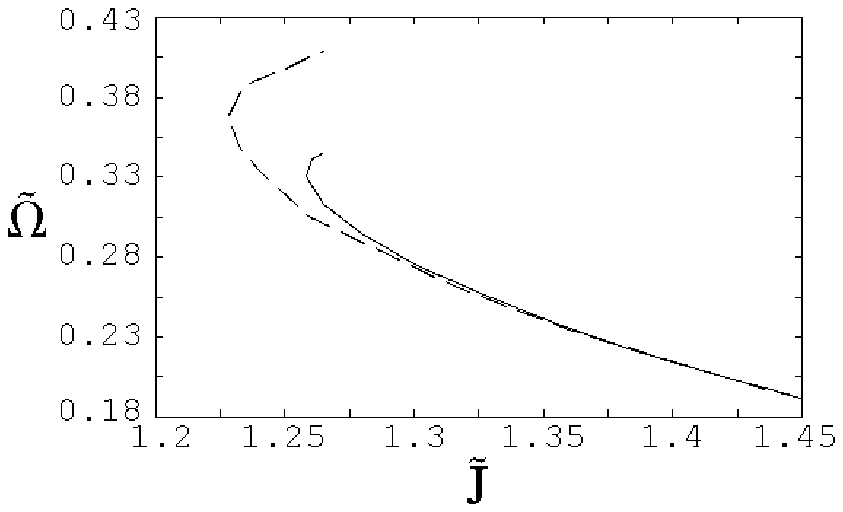}
\vspace{0.1cm}
\caption{Same as figure 1 but for the results of LRS1 and our present results 
with the $N=0.5$ equation of state.  }
\label{fig-N=05}
\end{figure}

\begin{figure}
\epsfbox{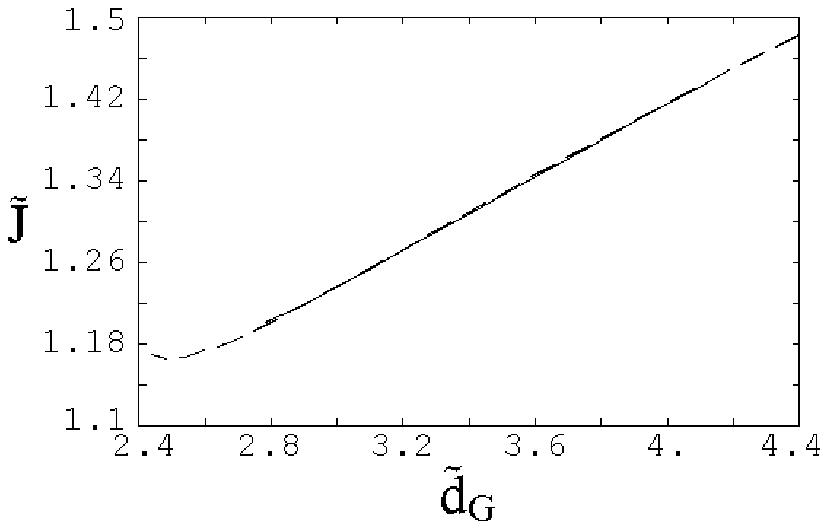}
\vspace{0.3cm}
\epsfbox{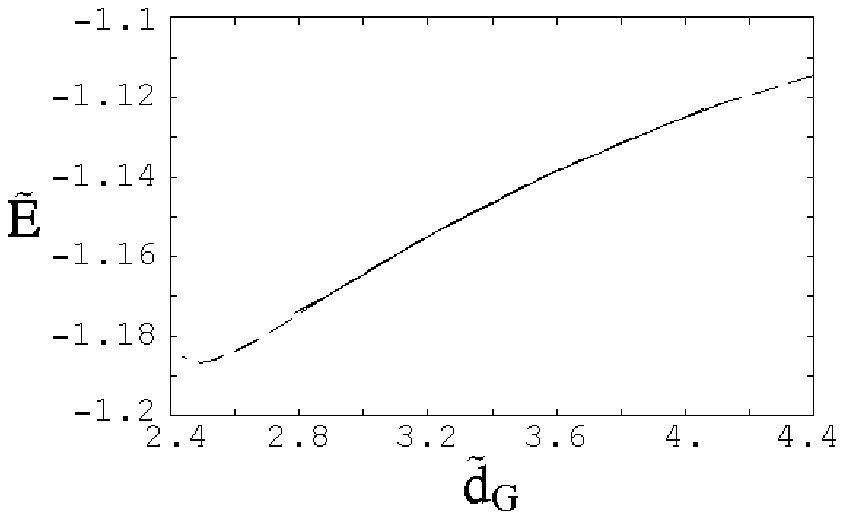}
\vspace{0.3cm}
\epsfbox{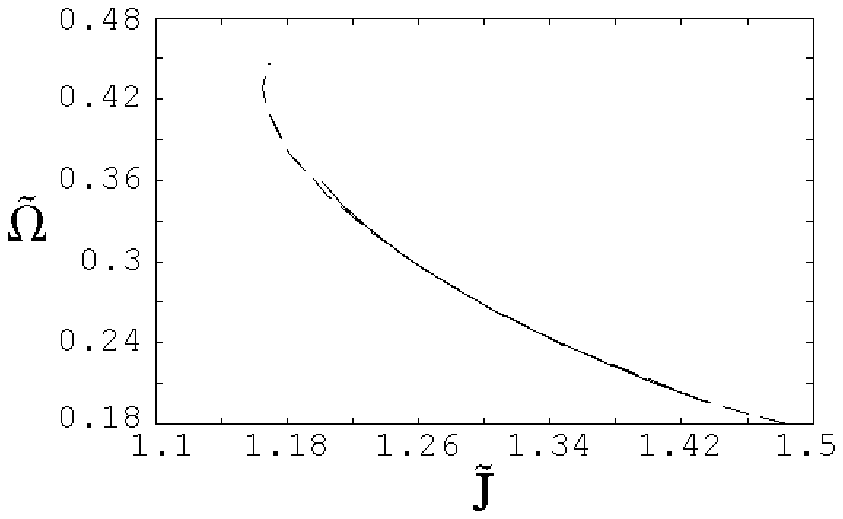}
\vspace{0.1cm}
\caption{Same as figure 2 but for the results with the $N=1$ 
equation of state.  }
\label{fig-N=1}
\end{figure}

\begin{figure}
\epsfbox{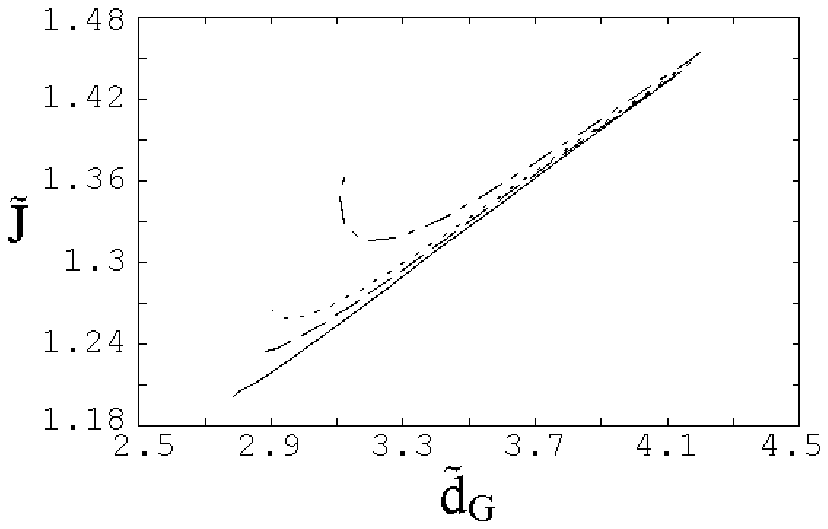}
\vspace{0.3cm}
\epsfbox{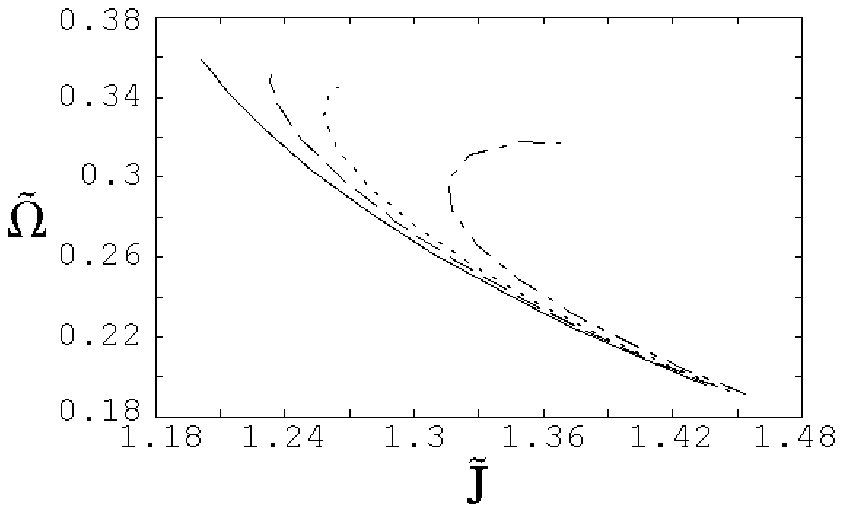}
\vspace{0.1cm}
\caption{Physical quantities of irrotational binary sequences for several 
polytropic indices.  (a) Total angular momentum as a function of a binary 
separation. (b) Orbital angular velocity as a function of the total angular 
momentum.  Different curves correspond to different polytropic indices:
$N=0$ (dash dotted line), $N=0.5$ (dotted line), $N=0.7$ (dashed line) 
and $N=1$ (solid line).  }
\label{fig-all}
\end{figure}

\section{Evolutionary sequences}

We have computed several sequences along each of which the mass, the
entropy ($K$) and the circulation (zero) are kept constant.  Thus these 
sequences can be regarded as the quasi-statically evolving sequences of 
inspiraling binary neutron stars due to gravitational wave emission.  
In figures 1, 2 and 3 we compare our results with those of the ellipsoidal 
approximation of the irrotational compressible binary systems of LRS1.  
The physical quantities of stationary sequences of 
irrotational equal mass binary systems with $N=0, 0.5$ and $1$ are shown in 
figures 1, 2 and 3, respectively.  In these figures, terminal points
at smaller separations or smaller angular momenta correspond
to models at contact stages of binary stars. The normalized values for
the total angular momentum $J$, the total energy $E$, the angular velocity 
$\Omega$ and a separation between the mass centers of two component 
stars $d_G$ are defined as follows, 
%
\[
{\tilde J}\,=\,{J \over (GM^3R_0)^{1/2}}\, ,\quad 
{\tilde E}\,=\,{E \over GM^2/R_0}\, ,\nonumber
\]
\begin{equation} \label{norm2}
{\tilde \Omega}\,=\,{\Omega \over (\pi G\bar\rho_0)^{1/2}}\quad 
{\rm and} \quad {\tilde d}_G\,=\,{d_G \over R_0} \, ,
\end{equation}
%
where $G$ is the gravitational constant, $M$ is the mass of one component 
star, $R_0$ is the radius of the spherical star with the same mass $M$ 
and the same polytropic index $N$. The quantity $\bar\rho_0$ is defined as 
$\bar\rho_0\,=\,M/(4\pi R^3_0/3)$.  This normalization is the same as 
that of LRS1.  Results are in good agreement with each other for models
with larger separations.  Differences at the separation $\ga 3.5$ are at worst 
$\la 0.5\%$ for any quantity.  In figure 1 we also show the results 
computed by our new computational method for {\it incompressible} binary 
with internal flow \cite{ue97a}.  Relative error between them are 
less than $\sim0.5\%$ everywhere.  These figures ensure the accuracy of 
our present results since the two independent computational methods give 
the same results for $N=0$ case even near the contact phase where the stars 
are significantly deformed by tidal field.   

On the other hand, for smaller separations, differences between LRS and 
our results become evident. For ellipsoidal approximations, curves with 
$N \la 1.2$ 
have turning points where the total angular momentum and the total energy 
attain their minimum values. In our irrotational binary models, curves turn 
only for nearly incompressible equations of state, but curves behave 
monotonically for compressible equations of state.  Since gravitational 
radiation carries away the angular momentum and the energy, the turning 
points on the curves correspond to critical points where dynamical 
instability sets in.  In this sense, our present results differ from those 
of LRS1 and LRS2. 

From the observational point of view, values of the angular velocity
have an important meaning because they are related to frequencies of 
emitted gravitational wave. In particular, those at critical points 
divide the ordered frequencies for quasi-periodic stages and the 
frequencies for dynamically inspiraling stages.  
As seen from figures 1 and 2 the critical frequencies obtained from the
ellipsoidal approximations are several percent larger that exact values
of our present irrotational binary models. 

These tendencies can be more explicitly shown in figure 4.
In figure 4, stationary sequences along which the mass, the entropy and the
circulation are the same, i.e. quasi-evolutionary sequences, are shown 
for polytropic indices $N=0$, $0.5$, $0.7$ and $1$. As seen from figure 4, 
evolutionary sequences with stiffer equations of state have turning
points but there are no turning points for more compressible 
equations of state.  As discussed above, models corresponding to turning 
points are critical ones beyond which dynamical instability of the binary 
stars will set in.  As figure 4 shows, for the binary stars with index 
$N \ga 0.7$, there are no critical points where dynamical 
instability will set in before contact stages.  It is important to note that 
the equation of state of neutron star matter can be approximately represented
by polytropes with index $N=0.5 \sim 1$ (see e.g. Shapiro \& Teukolsky 1983).  
Thus, as the separation decreases due to gravitational wave emission, 
the inviscid NS-NS binary will {\it stably} evolve to a contact configuration. 

\section{Discussion and conclusion}

In order to investigate evolutions of binary stars, several critical 
states should be taken into consideration. They are (1) a dynamically unstable 
state, (2) a Darwin-Riemann limit where the mass overflow from one component 
to the other will set in, and (3) a contact phase of binary stars.  
There can be varieties about the final fates of the binary neutron star
 system depending on the order of appearance of these critical states 
as the separation decreases due to gravitational wave emission.  
In particular, for the equal mass binary system, states of the contact phase 
and of onset of dynamical instability are expected to exist.  Our results show 
that for the binary system with $N \ga 0.7$ the dynamical instability point 
does never appear on the binary sequence but that binary stars do contact 
before dynamical instability sets in.  If merging after a contact phase 
proceeds quasi-statically, a dumbbell-like configuration will be 
formed \cite{eh85,ha86}.  Since this configuration will continue to evolve 
due to gravitational wave emission, it will become dynamically unstable 
at a certain point on this dumbbell sequence and begin to collapse 
violently because it is highly likely that there is a turning point on the 
dumbbell sequence.  

Although our analysis has been made in the framework of Newtonian gravity, 
there is a possibility that the same situation will occur even for general 
relativistic treatments.  If it will be the case for general relativity,
the scenario of evolution of irrotational binary systems should be changed.  
In particular, we should consider what 
will happen after contact phases of binary stars.  As for non-equal mass 
binary systems, there is a possibility that a similar situation will happen,
i.e. that the Darwin-Riemann limit or the contact phase will occur prior to 
dynamical instability.  

Consequently, for further quantitative research on real binary neutron star 
systems, it is important to take into account the general relativistic effect. 
General relativity is apt to make the radius smaller and the density 
distribution to more centrally condensed.  The typical relativistic
effect is the existence of the innermost stable circular orbit
(see e.g. Lincoln \& Will 1990; Kidder, Will \& Wiseman 1992).
However, as Lai, Rasio \& Shapiro~(1993b) discussed, the radius of this 
orbit is well inside the radius of the onset of dynamical instability due to 
the hydrodynamical (tidal) effect for most binary neutron star systems as 
far as the quantity $R\,c^2/GM$ satisfies the relation $Rc^2/GM\ga5$, where 
$R$ and $c$ are the stellar radius and the speed of light, respectively.  
This implies that for irrotational binary neutron star systems contact phases 
will appear first as the separation decreases.  In order to investigate 
general relativistic effects on the fate of binary neutron stars 
quantitatively, the competition among the tidal effect to elongate the stellar 
size and the relativistic effect to make it compact will be required.  
Recently the synchronously rotating NS-NS binary systems in general relativity 
have been approximately computed by several authors (Shibata 1994, 1997; 
Baumgarte et al. 1997).  We may be able to extend our 
method of calculating irrotational binary configuration in the framework 
of general relativity by using a similar formulation as the present
Letter.  

The advanced gravitational wave observatory mentioned in Introduction will 
certainly detect gravitational waves from the final phase of NS-NS binary 
systems in the early years of the next century.  In order to analyze 
gravitational waves and get information from these events, it is necessary 
to compute reliable wave forms at the merging stages.  Shibata, Nakamura 
\& Oohara (1992) performed dynamical computations of coalescing binary 
neutron stars and reported that the wave form depends on whether the 
component stars have spins or not (see also Rasio \& Shapiro 1996, and 
references therein).  
Therefore the spin effect plays an important role for the investigation of
determination of the equation of state for neutron star matter 
by using the wave form evaluated from hydrodynamical simulations.  
However, in these dynamical computations they began their computations from 
initial states which were not in exact equilibrium of binary stars 
with spins.  
In order to have more reliable results, it would be desirable to begin 
dynamical computations from exact equilibrium states of spinning binary 
stars.  
Our new solutions in this Letter can be used as initial states for such 
dynamical computations of coalescing binary stars.

\section*{Acknowledgments}

We would like to thank Dr. Shin'ichiro Yoshida for discussions.  
We also would like to thank Prof. Stuart L. Shapiro for correcting 
some misstatements in the manuscript.  
One of us (KU) would like to thank Profs. Dennis W. Sciama and 
John C. Miller and Dr. Antonio Lanza for their warm hospitality at 
ICTP and SISSA.  He would also like to thank Prof. Marek Abramowicz and 
Dr. Vladimir Karas for their encouragements. A part of numerical 
computations was carried out at the Astronomical Data Analysis Center of 
the National Astronomical Observatory, Japan.


\begin{thebibliography}{}

\bibitem[\protect\citename{Abramovici et al.\ }1992]{ab92} 
  Abramovici A. et al.,1992, Science, 256, 325
\bibitem[\protect\citename{Baumgarte et al.\ }1997]{ba97} 
  Baumgarte T. W. et al., 1997, Phys. Rev. Lett., submitted 
  (gr-qc/9704024/9705023)
\bibitem[\protect\citename{Bildsten \& Cutler\ }1992]{bc92} 
  Bildsten L., Cutler C., 1992, ApJ, 400, 175
\bibitem[\protect\citename{Cutler et al.\ }1993]{cu93} 
  Cutler C. et al., 1993, Phys. Rev. Lett., 70, 2984
\bibitem[\protect\citename{Eriguchi \& Hachisu\ }1985]{eh85} 
  Eriguchi Y., Hachisu I., 1985, A\&A, 142, 256
\bibitem[\protect\citename{Hachisu\ }1986]{ha86} 
  Hachisu I., 1986, ApJS, 62, 461
\bibitem[\protect\citename{Hachisu \& Eriguchi\ }1984]{he84} 
  Hachisu I., Eriguchi Y., 1984, Publ. Astron. Soc. Japan, 36, 259
\bibitem[\protect\citename{Kidder, Will \& Wiseman\ }1992]{ki92} 
  Kidder L. E., Will C. M., Wiseman A. G., 1992, Class. Quantum Gravity,
  9, L125
\bibitem[\protect\citename{Kochanek\ }1992]{ko92} 
  Kochanek C. S., 1992, ApJ, 398, 234
\bibitem[\protect\citename{Lai, Rasio \& Shapiro\ }1993]{lrs93a} 
  Lai D., Rasio F. A., Shapiro S. L., 1993a, ApJS,  88, 205
\bibitem[\protect\citename{Lai, Rasio \& Shapiro\ }1993]{lrs93b} 
  Lai D., Rasio F. A., Shapiro S. L., 1993b, ApJ,  406, L63
\bibitem[\protect\citename{Lai, Rasio \& Shapiro\ }1994]{lrs94a} 
  Lai D., Rasio F. A., Shapiro S. L., 1994a, ApJ, 420, 811
\bibitem[\protect\citename{Lai, Rasio \& Shapiro\ }1994]{lrs94b} 
  Lai D., Rasio F. A., Shapiro S. L., 1994b, ApJ, 423, 344
\bibitem[\protect\citename{Lincoln \& Will\ }1990]{li90} 
  Lincoln C. W., Will C. M., 1990, Phys. Rev. D, 42, 1123
\bibitem[\protect\citename{Miller\ }1974]{mi74} 
  Miller B.D., 1974, ApJ, 187, 609
\bibitem[\protect\citename{Rasio \& Shapiro\ }1996]{rs96}
  Rasio F. A., Shapiro S. L., 1996, in Compact Stars in Binaries, 
  proceedings of IAU Symposium 165, eds. 
  van Paradijs J., van den Heuvel E. P. J., Kuulkers E.,
  Dordrecht, Kluwer Academic Publishers
\bibitem[\protect\citename{Shibata\ }1994]{sh94} 
  Shibata M., 1994, Prog. Theor. Phys., 91, 871
\bibitem[\protect\citename{Shibata\ }1997]{sh97} 
  Shibata M., 1997, Phys. Rev. D, 55, 6019
\bibitem[\protect\citename{Shibata, Nakamura \& Oohara\ }1992]{sno92} 
  Shibata M., Nakamura. T., Oohara K., 1992, Prog. Theor. Phys., 88, 1079
\bibitem[\protect\citename{Thorne }1994]{th94}
  Thorne K. S., 1994, in Sasaki M., ed, ``Relativistic Cosmology" 
  Proceedings of the 8-th Nishinomiya-Yukawa Memorial Symposium, 
  Universal academy press, Tokyo, p. 67
\bibitem[\protect\citename{Shapiro \& Teukolsky\ }1983]{st83}
  Shapiro S. L., Teukolsky S. A., 1983, 
  Black Holes, White Dwarfs and Neutron Stars, Wiley, New York
\bibitem[\protect\citename{Ury\=u \& Eriguchi\ }1996]{ue96}  
  Ury\=u K., Eriguchi Y., 1996, MNRAS, 282, 653 
\bibitem[\protect\citename{Ury\=u \& Eriguchi\ }1997a]{ue97a}  
  Ury\=u K., Eriguchi Y., 1997a, in preparation
\bibitem[\protect\citename{Ury\=u \& Eriguchi\ }1997b]{ue97b}  
  Ury\=u K., Eriguchi Y., 1997b, in preparation
\end{thebibliography}
\end{document}